\pdfoutput=1
\documentclass[%
reprint,
superscriptaddress,
amsmath,amssymb,
aps,
pre,
]{revtex4-2}

\expandafter\def\expandafter\normalsize\expandafter{%
  \normalsize  
  \setlength\abovedisplayskip{1.5ex} 
  \setlength\belowdisplayskip{2.2ex} 
  \setlength\abovedisplayshortskip{0.3ex} 
  \setlength\belowdisplayshortskip{1.5ex} 
  }

\usepackage{graphicx}
\usepackage{dcolumn}
\usepackage{bm}
\usepackage{xcolor} 
\usepackage{hyperref,xcolor} 
\hypersetup{colorlinks=true,allcolors=blue}
\usepackage{tikz}

\def \initg{\tikz[baseline=-0.5ex] {
\fill (0,0) circle (1.5pt) coordinate (i);
\fill (2ex,0) circle (1.5pt) coordinate (j);
\draw (i)--(j);}}

\begin{document}

\title{Divergence asymmetry and connected components \\ in a general duplication-divergence graph model}

\author{Dario Borrelli}

\email{dario.borrelli@unina.it}
\affiliation{University of Naples Federico II,\\ Theoretical Physics Div., I-80125, Naples, Italy 
}%

\date{\today}

\begin{abstract}
This Letter introduces a generalization of known duplication-divergence models for growing random graphs. This general duplication-divergence model includes a new coupled \textit{divergence asymmetry} rate, which allows to obtain the structure of random growing networks by duplication-divergence in a continuous range of configurations between two known limit cases: (i) \textit{complete asymmetric divergence}, i.e., divergence rates affect only edges of either the original or the copy vertex, and (ii) \textit{symmetric divergence}, i.e., divergence rates affect equiprobably both the original and the copy vertex. Multiple connected sub-graphs (of order greater than one) emerge as the divergence asymmetry rate slightly moves from the complete asymmetric divergence case.  Mean-field results of priorly published models are nicely reproduced by this generalization. In special cases, the connected components size distribution $C_s$ suggests a power-law scaling of the form $C_s \sim s^{-\lambda}$ for $s>1$, e.g., with $\lambda \approx 5/3$ for divergence rate $\delta \approx 0.7$.
\end{abstract}


\maketitle


How does the structure of networks emerge? What are the principles underlying network evolution that led to observed network structures? Sequentially growing network models have been paradigmatic in tackling this kind of questions \citep{krapivsky2001organization,albert2002statistical,dorogovtsev2008critical}. 

Among sequentially growing network models, duplication models are based on the principle of duplication of existing patterns of linkage among vertices \cite{sole2002model,kim2002infinite,vazquez2003growing,krapivsky2005network}. The duplication-divergence principle, in particular, is inspired by a theory of genome evolution \cite{ohno1970evolution}, thus, these models are particularly interesting for the understanding of the structure of biological networks like protein interaction networks. Duplication models are also of a broader interest, which includes any kind of growing network that may be based on copying mechanisms of existing patterns of linkage among vertices (e.g.,  scientific citation graphs \cite{newman2018networks}, world-wide-web graphs \cite{kumar2000stochastic}, online social graphs \cite{bhat2016densification}). 
 
Duplication models emerge besides the widely studied growing network model known as preferential-attachment \cite{dorogovtsev2000structure,vazquez2003growing}, i.e., vertices with more interactions tend to attract even more interactions (with either a linear or a non-linear attachment rate) by new vertices that join the network \cite{barabasi1999mean,krapivsky2001organization}. Instead, in duplication models, vertices to be duplicated (along with their edges) are typically chosen uniformly at random. Duplication models have indirectly shown effective preferential-attachment \cite{vazquez2003growing}, therefore they are among candidate principles for the emergence of preferential-attachment \cite{dorogovtsev2022nature}. 

An iteration of a discrete time duplication-divergence model consists of \textit{duplication} by a random uniform choice of an existing vertex duplicated into a copy vertex (with the same edges), and \textit{divergence}, i.e., probabilistic loss of duplicate edges \cite{ispolatov2005duplication}. A general duplication model is known as duplication-divergence-dimerization-mutation model \cite{cai2015mean}, in which divergence is accompanied by addition of new edges between the copy vertex and other vertices (mutation), and between the copy vertex and its original vertex (dimerization); deletion of vertices is also considered in prior duplication models \cite{farid2006evolving}.  The relevance of these fascinating models has been especially revealing in the context of biological networks \cite{kim2002infinite,pastor2003evolving,schneider2011modeling}: prior research has shown structural similarities with protein-protein interaction networks of different reference species \cite{sole2002model,vazquez2003modeling,pastor2003evolving}.  Particular attention is paid toward duplication-divergence models where no links are added except from those resulting from duplication, hence, the growing structure of resulting networks emerge purely from reuse of linkage patterns of randomly chosen vertices \cite{ispolatov2005cliques,ispolatov2005duplication}. 

\begin{figure}[!t]
\hspace{-0.1in}
\includegraphics[width=0.43\textwidth]{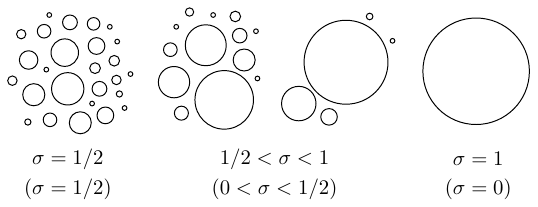}
\caption{Simplified depiction illustrating that the divergence asymmetry rate $\sigma \in [0,1]$ results in growing graphs with multiple connected components (of size $s>1$) as it slightly shifts away from the known complete asymmetric divergence case, i.e., $\sigma=1$ (or, $\sigma=0$ by symmetry). The known coupled symmetric divergence rate is $\sigma = 1/2$.}
\label{fig0}
\end{figure}

The \textit{divergence} process has typically interested only edges of the copy vertex, leaving intact the edges of the original vertex \cite{pastor2003evolving,ispolatov2005duplication}. This \textit{complete asymmetric divergence} generates graphs with a single connected component \cite{ispolatov2005duplication}, and possibly, vertices with no edges (hereafter called \textit{non-interacting vertices}). \textit{Symmetric divergence}, instead, is defined here as a divergence process that allows removal of a duplicate edge with same probability from both the copy vertex and the original vertex. Symmetric divergence can be coupled \cite{vazquez2003growing}, meaning that, given a duplicate edge, its removal can happen either from the original vertex or from the copy vertex (non-overlapping events), or uncoupled \cite{ispolatov2005cliques,sudbrack2018master}, where both the original and the copy vertex can independently lose the same duplicate edge. Differently from models with complete asymmetric divergence, models with symmetric divergence can exhibit connected components of heterogeneous size \cite{vazquez2003modeling,vazquez2003growing,ispolatov2005cliques}, and this feature is intriguing for graphs formed by connected components as well as for their interplay with percolation \cite{newman2007component,coniglio1982cluster,kim2002infinite}.

Albeit coupled symmetric divergence has been included in published models \cite{sole2002model, vazquez2003growing, ispolatov2005cliques, sole2008spontaneous}, here, for the first time, and unlike prior models, a general duplication-divergence model is introduced to encompass not only the complete asymmetric divergence and the coupled symmetric divergence cases, but also continuous extents of asymmetries in modeling divergence  (see Fig. \ref{fig0}). These divergence asymmetries allow graphs resulting from this model to be composed of multiple connected components of various sizes, in contrast with a special case of the model here introduced that recovers a known duplication-divergence model with complete asymmetric divergence, whose structure exhibits one connected component plus non-interacting vertices. In the following, for the general duplication-divergence model, relevant structural features like the mean number of edges, the mean vertex degree, vertex degree distribution, and connected components size distribution are studied, providing analytical and numerical results that emerge from new quantities introduced by this generalization.

\begin{figure}[!t]
\includegraphics[width=0.46\textwidth]{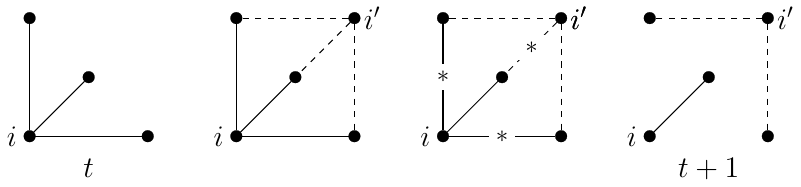}
\caption{At $t$, random uniform choice of vertex $i$ duplicated with all its edges (solid segments) into vertex $i'$; having $\sigma$ and $\delta$ $\neq 0,1$ yields possible complementary loss of duplicate edges (marked with $*$), resulting into the graph at $t+1$ with two connected components. Dashed segments are copy edges. }
\label{fig1}
\end{figure}

An undirected graph growing through a duplication-divergence network growth model is denoted here by $G_t=(N_t,E_t)$, where $N_t$ and $E_t$ are, respectively, the set of vertices and the set of edges at time $t$ of graph $G_t$. To avoid redundant notation, $N_t$ and $E_t$ also denote the number of vertices and the number of edges in $G_t$, respectively. As in traditional prior work on sequentially growing network models, in principle, time is considered a discrete variable as to have $N_t$ that increases by 1 at each iteration $t$ of the evolution process \footnote{This process may remind a single-gene duplication evolution; multiple genes duplications events (e.g., entire genome duplication) may be possible but they are not considered here in favor of a minimal approach.}. Hence, unless otherwise specified, the time variable $t$ equals $N_t$, and the growth process starts at $t_0 = N_{t_0}$ with $N_{t_0}$ vertices, possibly connected. A time scale separation between duplication and divergence events is assumed, so that divergence happens as soon as a duplication event occurs but before the subsequent duplication event. Such a time scale separation supports the idea that divergence occurs shortly after duplication events. At each $t$, duplication results in two exact copies, vertex $i$ and $i'$, of a randomly chosen vertex $i$, meaning that both $i$ and $i'$ have the same set of adjacent vertices $j$. Then, divergence changes this configuration by partially conserving duplicate edges. In particular, complementary preservation of duplicate edges allows divergence to conserve the edges of vertex $i$ by complementarily distributing them among vertices $i$ and $i'$  \cite{force1999preservation}. This process translates into a local broken symmetry: i.e., for each duplicate edge pair $\lbrace(i,j),(i',j)\rbrace$ only one of the two edges is conserved, either from $i$, with probability $\sigma$, or from $i'$, with probability $1-\sigma$. The probability $\sigma$ in this model is what is introduced here as the \textit{divergence asymmetry} rate; $\sigma$ allows to cover two limit cases: when $\sigma=\frac{1}{2}$ it is likely that vertices $i$ and $i'$ will lose, on average, the same number of edges in the divergence process, and this situation reflects the \textit{symmetric divergence} case. Conversely, when $\sigma = 1$ ($\sigma = 0$),  only vertex $i'$ (vertex $i$) will lose edges due to divergence, while vertex $i$ (vertex $i'$) conserves all of its edges; the latter situation reflects the \textit{complete asymmetric divergence}. When $\sigma = 1$ the model reduces to the complete asymmetric divergence case that has been studied in priorly published papers (see below). 

Besides the duplication and divergence principles, two additional sophistications are included in this generalization: \textit{dimerization} was introduced in prior research to allow interaction between the copy vertex and the original vertex \cite{vazquez2003modeling}; \textit{mutation} was also introduced in previous research to mimic the addition of new edges between the copy vertex and all other vertices except from $i$ and its adjacent vertices $j$ \cite{sole2002model}. Both dimerization and mutation add new edges besides edges that link vertex $i$ to adjacent vertices $j$ that are duplicated. 

The sequentially growing graph process is formalized by the following procedure occurring at a generic iteration $t$ (see also a depiction of a duplication-divergence (a)-(b) iteration in Fig. \ref{fig1}):

\begin{enumerate}
\item[(a)] \textit{Duplication}: a vertex $i$, chosen uniformly at random among interacting vertices with probability $d$, and among all vertices (including non-interacting ones) with probability $1-d$, is duplicated into a vertex $i'$ having the same edges of vertex $i$.

\item [(b)] \textit{Divergence}: for each pair of duplicate edges $\lbrace (i,j), (i',j) \rbrace$ linking $i$ and $i'$ to the same adjacent vertex $j$, only one of the two of edges of the pair is lost with probability $\delta$, either from vertex $i$ with probability $\sigma$, or from vertex $i'$ with probability $1-\sigma$.

\item[(c)] \textit{Dimerization}: one edge $(i,i')$ is added with probability $\alpha$ to link duplicate vertices.

\item[(d)] \textit{Mutation}: edges between the copy vertex $i'$ and all other vertices (except $i$ and its initial adjacent vertices) are added each with probability $\beta$.
\end{enumerate}

In agreement with prior work, the probability $\delta \in [0,1]$ is referred to as the \textit{divergence rate}; $\alpha \in [0,1]$ is called the \textit{dimerization rate}; $\beta \in [0,1]$ is called the \textit{mutation rate}. Here, we will consider $d=1$ and $d=0$, and mainly $\beta=\alpha= 0$, unless otherwise specified. As anticipated, this general duplication-divergence model generalizes the following known models: for $\sigma=\frac{1}{2}$ and $\alpha =0$, the growing process is the same as the one introduced in Ref. \cite{vazquez2003modeling} (without any addition of $(i,i')$ edge), while for $\sigma = 1$ and $\alpha=0$, the model reduces to the model in Ref. \cite{ispolatov2005duplication}. Differently from \cite{ispolatov2005duplication}, here, having a non-interacting vertex as a result of duplication-divergence is an allowed possibility and it occurs with probability $(1-\sigma)^k\delta^k + \sigma^k\delta^k$ for a vertex $i$ with $k$ adjacent vertices, while in Ref. \cite{ispolatov2005duplication} with probability $\delta^k$ a non-interacting vertex that results from duplication-divergence is removed from the graph. 


Firstly, the mean-field number of edges and mean vertex degree of $G_t$ with $d=0$ are calculated; here, $\alpha=\beta=0$ is set to facilitate readability (cases with $\alpha,\beta \neq 0$ are reported in Appendix \ref{app_A} and \ref{app_B}). The following recurrence equation can be written for the mean number of edges

\begin{equation}
	\langle E_{t+1} \rangle - \langle E_{t} \rangle = 2 \frac{\langle E_t \rangle}{t} - 2\delta \frac{\langle E_t \rangle}{t}.
	\label{E_t1}
\end{equation}

The gain term on the right hand side considers the duplication of $\langle k_t \rangle = 2 \langle E_t \rangle / t$ edges; the loss term considers a mean number of edges lost equal to

\begin{equation}
	\sigma \delta \frac{2\langle E_t \rangle}{t} + (1 - \sigma)\delta \frac{2 \langle E_t \rangle}{t}= 2 \delta \frac{\langle E_t \rangle }{t}.
\end{equation}

The exact solution to (\ref{E_t1}) for an initial graph with two connected vertices (i.e., $G_{t_0=2}$: \initg ) is

\begin{equation}
	\langle E_t \rangle = \frac{\Gamma(2-2\delta+t)}{\Gamma(2-2\delta + 2)\Gamma(t)},
	\label{ex_Et}
\end{equation}

with $\Gamma(\cdot)$ the Euler Gamma function. Having (\ref{ex_Et}), the mean vertex degree follows immediately from

\begin{equation}
	\langle k_t \rangle = 2\frac{\langle E_t \rangle}{t}.
	\label{avg_meandeg}
\end{equation}

To give a physical sense of the solution (\ref{ex_Et}), it is convenient to solve the continuum approximation of (\ref{E_t1})

\begin{equation}
	\frac{d\langle E_t \rangle}{dt} = \frac{2(1-\delta)}{t} \langle E_t \rangle,
\end{equation}

which returns the following scaling with $t$ for the number of edges
\begin{equation}
\langle E_t \rangle  \sim 
\begin{cases}
t^{2(1-\delta)}, \hspace{0.52cm}\mathrm{for \; \;} \delta \gtrless 1/2,\\
t, \hspace{1.4cm}\mathrm{for \; \;}\delta=1/2.
\end{cases}
\end{equation}

For the mean vertex degree, the scaling with $t$ is then

\begin{equation}
	\langle k_t \rangle  \sim 
\begin{cases}
t^{(1-2\delta)}, \hspace{0.52cm}\mathrm{for \; \;} \delta \gtrless 1/2,\\
\mathrm{const.}, \hspace{0.63cm}\mathrm{for \; \;}\delta=1/2.
\end{cases}
\label{avg_meandeg_d0}
\end{equation}

\begin{figure}
\includegraphics[width=0.43\textwidth]{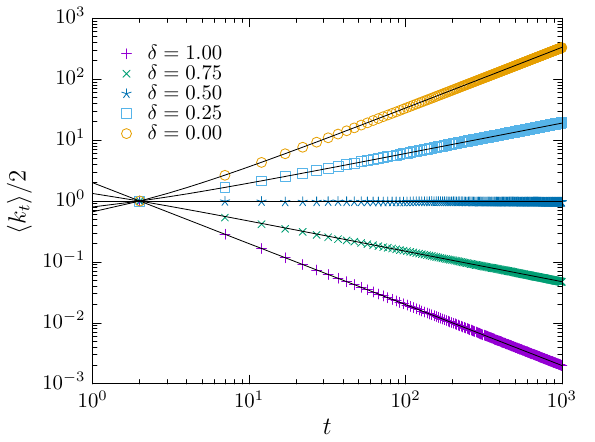}
\caption{Mean vertex degree for $\delta$ values (in legend) averaged over $10^3$ simulations of growing networks by the model with $d=0, \sigma=1/2, \alpha=\beta =0$. Each simulation starts with two connected vertices, and ends when the graph order is $10^3$ vertices; line-plots represent exact mean-field solution.}
\label{fig2}
\end{figure}

\begin{figure}
\includegraphics[width=0.43\textwidth]{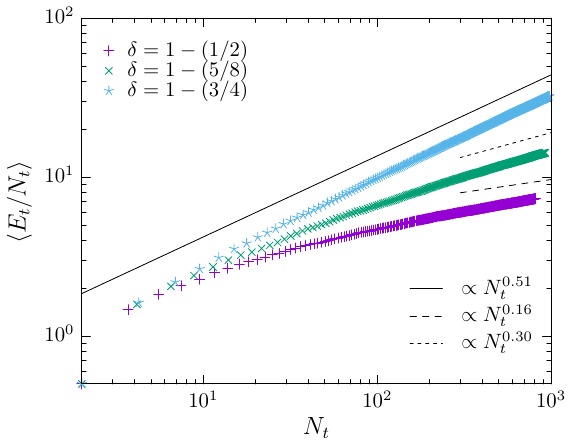}
\caption{$\langle E_t / N_t \rangle$ plotted versus $N_t$ (number of vertices with at least one edge) for $10^3$ simulations of the general duplication-divergence model with $d= \sigma=1, \alpha=\beta =0$, reproducing results of the model in \cite{ispolatov2005duplication}; $\delta$ values and predicted slopes (in legend with lines) are approached asymptotically.}
\label{fig2b}
\end{figure}

Fig.~\ref{fig2} plots the mean vertex degree (via (\ref{avg_meandeg})) versus numerical simulations of the model procedure with $d=\beta=\alpha=0$, and $\sigma=1/2$. For $d=\sigma=1$ (and $\beta=\alpha=0$), Fig.~\ref{fig2b} compares numerical simulations with results of the duplication-divergence model in \cite{ispolatov2005duplication}, in which non-interacting vertices were not considered (plotting the number of vertices with at least one edge $N_t$, since $t \neq N_t$ when $\delta \neq 0$ as in \cite{ispolatov2005duplication}). Concerning fluctuations about the mean number of edges, i.e., $\langle E_t^2 \rangle - \langle E_t \rangle^2$, the second moment $\langle E_t^2 \rangle$ is required. Following \cite{bhat2016densification}, for a single realization one writes the number of edges as

\begin{equation}
	E_{t+1} = E_t + v,
	\label{realiz_Et1}
\end{equation}

with $v$ a random variable in $[0,k]$ distributed as a binomial distribution 

\begin{equation}
	B(v|k) = \binom{k}{v}\delta^{k-v}(1-\delta)^v,
\end{equation}

having mean $\hat{v}=(1-\delta)k$, and second moment

\begin{equation}
	\hat{v}^2 = \sum_v v^2 B(v|k) = (1-\delta)^2k^2 +\delta(1-\delta)k.
\end{equation}

By squaring (\ref{realiz_Et1}) and averaging over the ensemble of realizations, one obtains

\begin{equation}
\begin{aligned}
\langle E_{t+1}^2 \rangle = \langle E_t^2 \rangle + (1-\delta)^2\langle k_t^2 \rangle + \\ + \delta(1-\delta)\langle k_t \rangle + (1-\delta)\frac{4\langle E_t^2 \rangle}{t}.
\end{aligned}
\label{2ndM_Et1}
\end{equation}

From this, one can get the fluctuation about the mean number of edges, which scales with $t$ as

\begin{equation}
	\langle E_t^2 \rangle - \langle E_t \rangle^2  \sim 
	\begin{cases}
t, \hspace{1.3cm}\mathrm{for \; \;} \delta > 3/4, \\
t \mathrm{ln}(t), \hspace{0.60cm}\mathrm{for \; \;}\delta=3/4, \\
t^{4-4\delta}, \hspace{0.64cm}\mathrm{for \; \;}0<\delta<3/4.
	\end{cases}
\end{equation}

Note that the above result is the same expression shown in Ref.~\cite{bhat2016densification}, whose copying model is recovered from this generalization when setting $d=0$, $\sigma=1$, $\alpha=1$, $\beta=0$.

\begin{figure}[!t]
\includegraphics[width=0.445\textwidth]{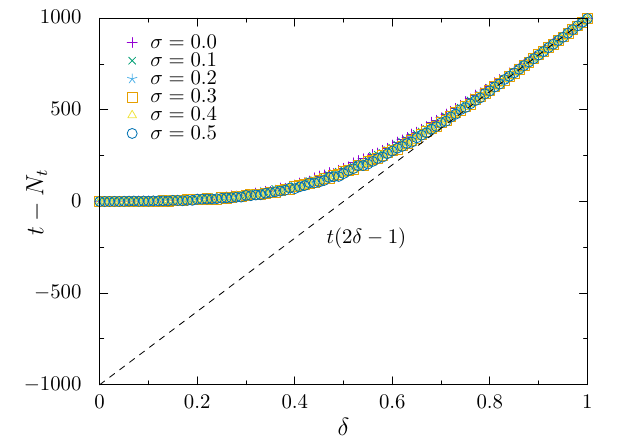}
\caption{Number of non-interacting vertices $t - N_t$ versus $\delta$ for various $\sigma$ (in legend), $t=10^3$. For $d=1$ and $\delta > 1/2$, $\eta = 2\delta -1$ approximates the rate of joining the set of non-interacting vertices (independent of $\sigma$);  $\mu = 1-\eta = 2(1-\delta)$ represents the rate of joining $N_t$ vertices with at least one edge.}
\label{fig_I0}
\end{figure}

For the vertex degree distribution, one can consider the expected number of vertices with degree $k$ at time $t$, denoted by $N_k(t):=\langle N_k(t) \rangle$, to write its rate of change $\partial N_k(t)/\partial t$. Knowing that $N_k(t) = t n_k(t)$ with $n_k(t)$ the fraction of vertices with degree $k$, it yields

\begin{equation}
	\frac{\partial N_k(t)}{\partial t} = t \frac{\partial n_k(t)}{\partial t} + n_k(t).
\label{starting_lhs}
\end{equation}
\\
With a stationary vertex degree distribution $n_k(t)=n_k$, for any $t'>t$, one gets

\begin{equation}
	\frac{d N_k}{d t} = n_k.
\end{equation}

When $N_t \neq t$ but generically $N:=N_t=\mu t$, with $\mu$ a constant rate (with respect to time) of joining the set of vertices with degree $k\geq1$, then one can write

\begin{equation}
	\mu \frac{dN_k}{dN} = \mu n_k.
	\label{as_dnk}
\end{equation}

From these considerations, through a rate equation approach \cite{krapivsky2010kinetic}, an evolution equation for the vertex degree distribution can be written. The rate $\mu$ (similarly introduced in \cite{ispolatov2005duplication}) is defined as the rate at which vertex $i'$ conserves at least one edge after divergence; here, $\mu$ may depend on some parameters of the general model, and in particular, on the value of $d$. Then, the rate equation for the evolution of the number of $k$-degree vertices, $N_k$, is

\begin{equation}
	\mu \frac{dN_k}{dN} = (1-\delta) \left[ (k-1)n_{k-1}-kn_k\right] + \mathcal{M}_k^{\sigma} + \mathcal{M}_k^{1-\sigma},
	\label{masteq}
\end{equation}

which is reminiscent of similar rate equations introduced in  \cite{kim2002infinite, ispolatov2005duplication}, yet, the last two terms on the right hand side are respectively the following sum

\begin{equation}
	\mathcal{M}_k^{\sigma} = \sum_{s \geq k} \binom{s}{k}[\sigma (1-\delta)]^k [ 1 -\sigma (1-\delta)]^{s-k}n_s, 
	\label{m1}
\end{equation}

and 

\begin{equation}
	\mathcal{M}_k^{1 - \sigma} = \sum_{s \geq k} \binom{s}{k}[(1 - \sigma) (1-\delta)]^k [ 1 -(1 - \sigma) (1-\delta)]^{s-k}n_s.
	\label{m2}
\end{equation}

For $k \gg 1$, Eq.~(\ref{masteq}) is conveniently rewritten with a continuum approach

\begin{equation}
	\mu \frac{dN_k}{dN} + (1-\delta)\frac{d(n_k k)}{dk} = \mathcal{M}_k^{\sigma} + \mathcal{M}_k^{1-\sigma}. 
\label{masteq_rec}
\end{equation}

One can leverage on the result of \cite{kim2002infinite} to find an approximate form of the two terms on the right hand side of (\ref{masteq_rec}) as their summands are sharply peaked respectively around $s \approx k/\sigma (1-\delta)$, and $s \approx k/(1-\sigma)(1-\delta)$. The two terms become $M_k^{\sigma} \approx n_{k/\sigma(1-\delta)}[\sigma(1-\delta)]^{-1}$, and $M_k^{1-\sigma} \approx n_{k/(1-\sigma)(1-\delta)}[(1-\sigma)(1-\delta)]^{-1}$ (see \cite{kim2002infinite,ispolatov2005duplication} for a similar approach). Then, Eq.~(\ref{masteq_rec}) becomes

\begin{widetext}
\begin{equation} 
	\mu \frac{dN_k}{dN} + (1-\delta)\frac{d(n_k k)}{dk} = n_{k/\sigma(1-\delta)}[\sigma(1-\delta)]^{-1} + n_{k/(1-\sigma)(1-\delta)}[(1-\sigma)(1-\delta)]^{-1}. 
\label{masteq_rec2}
\end{equation}
\end{widetext}

As carried out priorly \cite{kim2002infinite,ispolatov2005duplication}, the above Eq.~(\ref{masteq_rec2}) is specialized for $\frac{1}{2} \leq \delta<1$ by using Eq.~(\ref{as_dnk}) and by assuming a power-law scaling $n_k \sim k^{-\gamma}$. From Eq.~(\ref{masteq_rec2}), one gets

\begin{equation}
\begin{aligned}
	 	& \mu  + (1-\delta)(1-\gamma) =  \\ & =(1-\delta)^{\gamma -1}[\sigma^{\gamma -1} + (1-\sigma)^{\gamma-1}].
\end{aligned}
\label{gammaexp}
\end{equation}

Eq.~(\ref{gammaexp}) generalizes prior findings concerning the exponent of the assumed power-law vertex degree distribution; indeed, one can notice (see Fig.~\ref{fig_I0}) that when $d=1$, the rate $\mu$ is independent of $\sigma$ and of the order of the growing graph, and also that, as $\delta$ increases, $\mu \rightarrow 2(1-\delta)$, which holds well for  $\delta > 1/2$. As numerically suggested in Fig.~\ref{fig_I0}, when $\delta > 1/2$, then $\mu = 1 - \eta$, being $\eta$ the rate of joining the set of non-interacting vertices, in agreement with the choice of $\mu$ in \cite{ispolatov2005duplication}. Then, with $\mu = 2(1-\delta)$, Eq.~(\ref{gammaexp}) gives 
\begin{equation}
	\gamma = 
	\begin{cases}
	3 - (1-\delta)^{\gamma -2}, \hspace{1cm}\mathrm{for \; \;} \sigma = 0,1,\\
	3 - 2^{2-\gamma}(1-\delta)^{\gamma-2}, \hspace{0.28cm}\mathrm{for \; \;} \sigma = 1/2, \\
	3 - g_{\gamma,\sigma}(1-\delta)^{\gamma -2}, \hspace{0.41cm}\mathrm{for \; \;} \sigma \gtrless 1/2,
	\end{cases}
	\label{isp_gener}
\end{equation}

\begin{figure}[!t]
\includegraphics[width=0.445\textwidth]{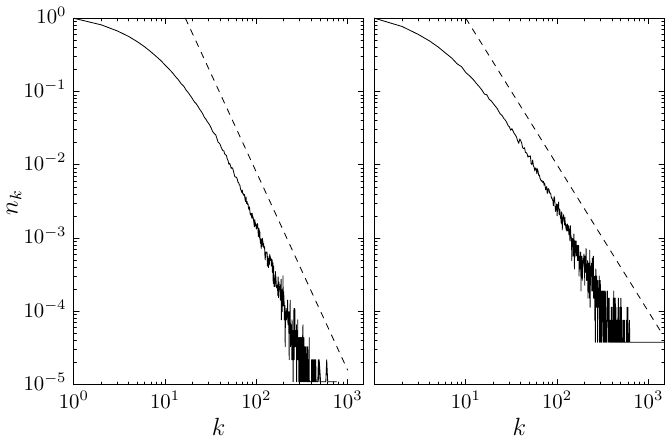}
\caption{Plots of $n_k$ for simulated graphs (solid curve) and power-laws for visual reference (dashed). Values of $n_k$ were scaled between 0 and 1 dividing by the maximum. Left panel: $n_k$ for model graphs with $t=10^7$ and $\delta=\sigma=1/2, d=\alpha=\beta=0$. Right panel: $n_k$ for model graphs with $t=10^6$ and $\delta=1/2,d=\sigma=1,\alpha=\beta=0$. Power-laws have exponents respectively of $\gamma \approx 2.5$ (left panel), and $\gamma \approx 2$ (right panel).} 
\label{fig4b}
\end{figure}

with $g_{\gamma,\sigma} = \sigma^{\gamma-1} + (1-\sigma)^{\gamma-1}$. Eq.~(\ref{isp_gener}) generalizes $\gamma$ introduced in \cite{ispolatov2005duplication}, which is precisely obtained by setting $\sigma=0$ (or, by symmetry, $\sigma=1$) and $d=1$,  recalling that $d=1$ results into a duplication event that choses a vertex $i$ among all vertices with at least one edge. 
Instead, when $d=0$, and $\mu$ is set equal to 1 in Eq.~(\ref{gammaexp}) (which is plausible for example if we assume $\alpha=1$ like in \cite{bhat2016densification}),  we get the following relations for the exponent $\gamma$
\vspace{-0.1in}

\begin{equation}
	\gamma = 
	\begin{cases}
	1 + \frac{1}{1-\delta} - (1-\delta)^{\gamma -2}, \hspace{0.96cm}\mathrm{for \; \;} \sigma = 0,1,\\
	1 +  \frac{1}{1-\delta} - 2^{2-\gamma}(1-\delta)^{\gamma-2}, \hspace{0.25cm}\mathrm{for \; \;} \sigma = 1/2, \\
	1 + \frac{1}{1-\delta} - g_{\gamma,\sigma}(1-\delta)^{\gamma -2}, \hspace{0.38cm}\mathrm{for \; \;} \sigma \gtrless 1/2.
	\end{cases}
	\label{kim_gener}
\end{equation}

\begin{figure}[!b]
\includegraphics[width=0.465\textwidth]{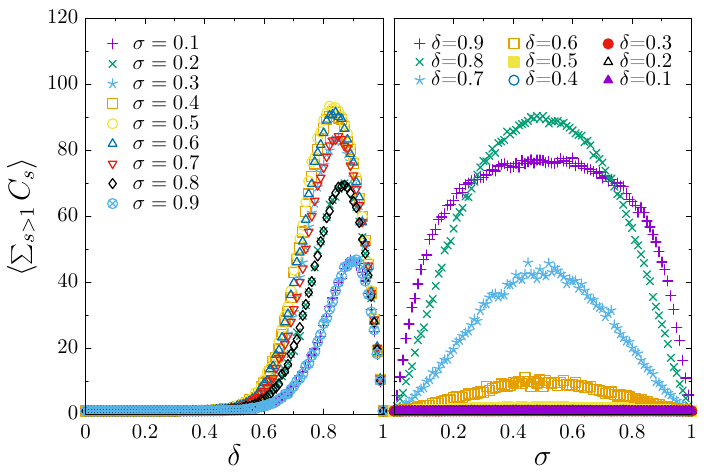}
\caption{Left panel: mean number of components of size $s>1$ as a function of $\delta$ for various $\sigma$ (in legend); $10^4$ simulations of the model ended at $t=10^3$ with parameters $d=1,\alpha=\beta=0$. Right panel: mean number of components of size $s>1$ as a function of $\sigma$ for various $\delta$ (in legend).} 
\label{fig6}
\end{figure}
 
Eq.~(\ref{kim_gener}) generalizes the expression for the exponent $\gamma$ introduced in \cite{kim2002infinite}, which is manifestly obtained when we set $\sigma=0$ (or, by symmetry, $\sigma=1$). 

Note that in duplication-divergence with $d=0$ (and $\alpha=\beta=0)$, the value of $\delta$ for which it may be plausible to consider a limiting power-law vertex degree distribution is when $\delta =1/2$, which follows directly from (\ref{avg_meandeg_d0}) having a constant average vertex degree. For $d=\alpha=\beta=0$, $\delta=\sigma=1/2$, one gets $\gamma=5/2$, which is in good agreement with simulations (Fig.~\ref{fig4b}). 

To obtain Eq.~(\ref{gammaexp}), a time-independent vertex degree distribution was assumed, since we have turned (\ref{starting_lhs}) into (\ref{as_dnk}) leading to (\ref{masteq_rec2}). If one considers a non-stationary time-dependent vertex degree distribution, the first term on the left hand side of (\ref{masteq_rec2}) would be the right hand side of (\ref{starting_lhs}). The resulting time-dependent form of the rate equation may not have a straightforward analtytic solution. Yet, in \cite{vazquez2003growing}, for a special case of the general model with $\sigma=1/2$ and $\beta=d=0$, moments of the vertex degree distribution were calculated, leading to the emergence of multifractality \cite{dorogovtsev2002multifractal}.

\begin{figure}[!t]
\includegraphics[width=0.43\textwidth]{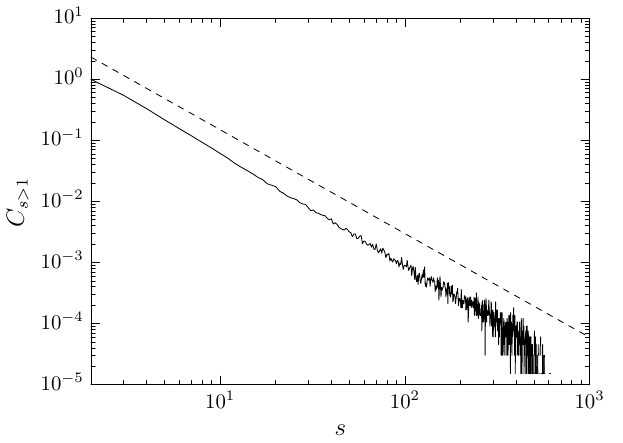}
\caption{$C_{s>1}$ for the duplication-divergence model with $d=1, \delta=0.7,\sigma=1/2, \alpha= \beta=0$, which is obtained from $10^2$ simulations ended when $t=5 \cdot 10^3$; dashed line is for visual reference of a power-law with exponent $-\lambda \approx - 5/3$.  Values of $C_{s>1}$ were scaled between 0 and 1 dividing by the maximum. As an intriguing note, the power-law exponent reminds that of $-5/3$ Kolmogorov isotropic turbulence, exponent that might firstly appeared in \cite{onsager1949statistical}.} 
\label{fig7}
\end{figure}

\begin{figure}[h!] 
\includegraphics[width=0.43\textwidth]{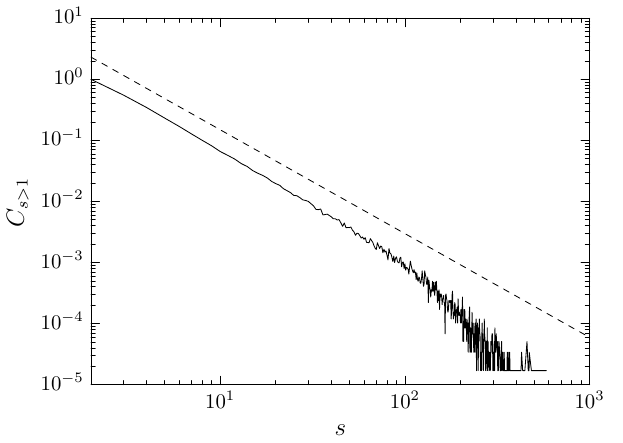}
\caption{$C_{s>1}$ for the model with $d=0,\delta=0.7,\sigma=1/2, \alpha= \beta=0$ that is obtained from $10^2$ simulations ended when $t=5 \cdot 10^3$; Values of $C_{s>1}$ were scaled between 0 and 1 dividing by the maximum. Dashed line is a power-law $C_{s>1} \sim s^{-\lambda}$, with $\lambda \approx  5/3$ as in the case of $d=0$ (Fig.~\ref{fig7}); noticing a faster decay (than in Fig.~\ref{fig7}) for larger values of $s$.} 
\label{fig8}
\end{figure}

As anticipated in the simplified depiction of Fig.~\ref{fig0}, the effect of having a divergence asymmetry rate $\sigma$ can be appreciated when computing the number of connected components as well as their size distribution $C_s$. Indeed, when varying $(\sigma,\delta,d)$, graphs grown by the general duplication-divergence model can exhibit multiple connected sub-graphs (i.e., components) of heterogeneous sizes for a continuous range of $\sigma$ values between complete asymmetric divergence ($\sigma=0$ or $\sigma=1$) and symmetric divergence ($\sigma = 1/2$).  

Fig.~\ref{fig6} (right panel) (with $d=1$) shows the mean number of connected components of size $s>1$, namely $\langle\sum_{s>1}C_s \rangle$, versus $\sigma$, when varying divergence rate $\delta$. As $\sigma$ departs from the complete asymmetric divergence case (i.e., when $\sigma \neq 0$ or $\sigma \neq 1$), the number of connected components $\langle\sum_{s>1}C_s \rangle$ increases, reaching a maximum at $\sigma = 1/2$ (symmetric divergence) for any value of $0<\delta <1$.  Similarly, in Fig.~\ref{fig6} (left panel), the number of connected components of size greater than 1 is plotted versus $\delta$ for diverse $\sigma$ values. As $\sigma \gtrless 1/2$, values of $\langle\sum_{s>1}C_s \rangle$ show overlap on top of each other (e.g., $\sigma=0.2$ and $\sigma = 0.8$ collapse on the same curve), which reflects the symmetric nature of $\sigma$ as well as it reflects that the original vertex and the copy vertex are indistinguishable in coupled divergence.  Then, for $\delta \in [0.6,0.9]$ curves exhibit a maximum number of connected components (of size greater than 1), with $\delta$ corresponding to a maximum with a shift toward higher $\delta$ values as $\sigma \rightarrow 0,1$ (Fig.~\ref{fig6}, left panel).  For values of $\delta \approx 0.7$, the expected proportion of connected components of size $s$, $C_{s>1}$, obtained numerically, shows power-law scaling $C_{s>1} \sim s^{-\lambda}$ with $\lambda \approx -5/3$ (see Fig.~\ref{fig7}). A similar power-law scaling with $\lambda \approx -5/3$ is shown in Fig.~\ref{fig8} for $d=0$, where a slightly faster decay emerges for large component sizes.

This Letter introduced a general model of random graph growth via duplication-divergence. As a main contribution, the divergence process includes continuous extent of asymmetry due to a newly introduced divergence asymmetry rate that yield diverse structural configurations among which those of prior models (namely, complete divergence asymmetry and divergence symmetry). The extent of divergence asymmetry can be responsible for the emergence of connected components of various sizes whose distribution may scale algebraically in special cases of the general model. This feature is very intriguing as many empirical networks (whose growth may be driven by duplication-divergence) have shown to exhibit connected components of heterogeneous size.  The mean-field number of edges and mean vertex degree calculated here show that their analytic form well generalizes prior results. In particular, the general asymptotic vertex degree distribution derived here, which is relevant in a plethora of studies on sparse network structures, allows to obtain well known exponents for the assumed power-law vertex degree distribution, generalizing their form with the here introduced variable -- divergence asymmetry rate $\sigma$. Concerning both the expected vertex degree distribution and connected components of size $s$, this Letter has limited the study (numerical for the connected component size distribution) to particular ranges of model parameters to emphasize discussed findings.

D.B. acknowledges seminars held by the Physics Dept. and by the Dieti Dept., at the University of Naples, which have been of inspiration for this research.

\bibliography{biblio}

\providecommand{\noopsort}[1]{}\providecommand{\singleletter}[1]{#1}%
\begin{thebibliography}{30}%
\makeatletter
\providecommand \@ifxundefined [1]{%
 \@ifx{#1\undefined}
}%
\providecommand \@ifnum [1]{%
 \ifnum #1\expandafter \@firstoftwo
 \else \expandafter \@secondoftwo
 \fi
}%
\providecommand \@ifx [1]{%
 \ifx #1\expandafter \@firstoftwo
 \else \expandafter \@secondoftwo
 \fi
}%
\providecommand \natexlab [1]{#1}%
\providecommand \enquote  [1]{``#1''}%
\providecommand \bibnamefont  [1]{#1}%
\providecommand \bibfnamefont [1]{#1}%
\providecommand \citenamefont [1]{#1}%
\providecommand \href@noop [0]{\@secondoftwo}%
\providecommand \href [0]{\begingroup \@sanitize@url \@href}%
\providecommand \@href[1]{\@@startlink{#1}\@@href}%
\providecommand \@@href[1]{\endgroup#1\@@endlink}%
\providecommand \@sanitize@url [0]{\catcode `\\12\catcode `\$12\catcode
  `\&12\catcode `\#12\catcode `\^12\catcode `\_12\catcode `\%12\relax}%
\providecommand \@@startlink[1]{}%
\providecommand \@@endlink[0]{}%
\providecommand \url  [0]{\begingroup\@sanitize@url \@url }%
\providecommand \@url [1]{\endgroup\@href {#1}{\urlprefix }}%
\providecommand \urlprefix  [0]{URL }%
\providecommand \Eprint [0]{\href }%
\providecommand \doibase [0]{https://doi.org/}%
\providecommand \selectlanguage [0]{\@gobble}%
\providecommand \bibinfo  [0]{\@secondoftwo}%
\providecommand \bibfield  [0]{\@secondoftwo}%
\providecommand \translation [1]{[#1]}%
\providecommand \BibitemOpen [0]{}%
\providecommand \bibitemStop [0]{}%
\providecommand \bibitemNoStop [0]{.\EOS\space}%
\providecommand \EOS [0]{\spacefactor3000\relax}%
\providecommand \BibitemShut  [1]{\csname bibitem#1\endcsname}%
\let\auto@bib@innerbib\@empty
\bibitem [{\citenamefont {Krapivsky}\ and\ \citenamefont
  {Redner}(2001)}]{krapivsky2001organization}%
  \BibitemOpen
  \bibfield  {author} {\bibinfo {author} {\bibfnamefont {P.~L.}\ \bibnamefont
  {Krapivsky}}\ and\ \bibinfo {author} {\bibfnamefont {S.}~\bibnamefont
  {Redner}},\ }\bibfield  {title} {\bibinfo {title} {Organization of growing
  random networks},\ }\href
  {https://doi.org/https://doi.org/10.1103/PhysRevE.63.066123} {\bibfield
  {journal} {\bibinfo  {journal} {Physical Review E}\ }\textbf {\bibinfo
  {volume} {63}},\ \bibinfo {pages} {066123} (\bibinfo {year}
  {2001})}\BibitemShut {NoStop}%
\bibitem [{\citenamefont {Albert}\ and\ \citenamefont
  {Barab{\'a}si}(2002)}]{albert2002statistical}%
  \BibitemOpen
  \bibfield  {author} {\bibinfo {author} {\bibfnamefont {R.}~\bibnamefont
  {Albert}}\ and\ \bibinfo {author} {\bibfnamefont {A.-L.}\ \bibnamefont
  {Barab{\'a}si}},\ }\bibfield  {title} {\bibinfo {title} {Statistical
  mechanics of complex networks},\ }\href
  {https://doi.org/https://doi.org/10.1103/RevModPhys.74.47} {\bibfield
  {journal} {\bibinfo  {journal} {Reviews of Modern Physics}\ }\textbf
  {\bibinfo {volume} {74}},\ \bibinfo {pages} {47} (\bibinfo {year}
  {2002})}\BibitemShut {NoStop}%
\bibitem [{\citenamefont {Dorogovtsev}\ \emph {et~al.}(2008)\citenamefont
  {Dorogovtsev}, \citenamefont {Goltsev},\ and\ \citenamefont
  {Mendes}}]{dorogovtsev2008critical}%
  \BibitemOpen
  \bibfield  {author} {\bibinfo {author} {\bibfnamefont {S.~N.}\ \bibnamefont
  {Dorogovtsev}}, \bibinfo {author} {\bibfnamefont {A.~V.}\ \bibnamefont
  {Goltsev}},\ and\ \bibinfo {author} {\bibfnamefont {J.~F.}\ \bibnamefont
  {Mendes}},\ }\bibfield  {title} {\bibinfo {title} {Critical phenomena in
  complex networks},\ }\href
  {https://doi.org/https://doi.org/10.1103/RevModPhys.80.1275} {\bibfield
  {journal} {\bibinfo  {journal} {Reviews of Modern Physics}\ }\textbf
  {\bibinfo {volume} {80}},\ \bibinfo {pages} {1275} (\bibinfo {year}
  {2008})}\BibitemShut {NoStop}%
\bibitem [{\citenamefont {Sol{\'e}}\ \emph {et~al.}(2002)\citenamefont
  {Sol{\'e}}, \citenamefont {Pastor-Satorras}, \citenamefont {Smith},\ and\
  \citenamefont {Kepler}}]{sole2002model}%
  \BibitemOpen
  \bibfield  {author} {\bibinfo {author} {\bibfnamefont {R.~V.}\ \bibnamefont
  {Sol{\'e}}}, \bibinfo {author} {\bibfnamefont {R.}~\bibnamefont
  {Pastor-Satorras}}, \bibinfo {author} {\bibfnamefont {E.}~\bibnamefont
  {Smith}},\ and\ \bibinfo {author} {\bibfnamefont {T.~B.}\ \bibnamefont
  {Kepler}},\ }\bibfield  {title} {\bibinfo {title} {A model of large-scale
  proteome evolution},\ }\href {https://doi.org/10.1515/9781400841356.396}
  {\bibfield  {journal} {\bibinfo  {journal} {Advances Complex Systems}\
  }\textbf {\bibinfo {volume} {5}},\ \bibinfo {pages} {43} (\bibinfo {year}
  {2002})}\BibitemShut {NoStop}%
\bibitem [{\citenamefont {Kim}\ \emph {et~al.}(2002)\citenamefont {Kim},
  \citenamefont {Krapivsky}, \citenamefont {Kahng},\ and\ \citenamefont
  {Redner}}]{kim2002infinite}%
  \BibitemOpen
  \bibfield  {author} {\bibinfo {author} {\bibfnamefont {J.}~\bibnamefont
  {Kim}}, \bibinfo {author} {\bibfnamefont {P.}~\bibnamefont {Krapivsky}},
  \bibinfo {author} {\bibfnamefont {B.}~\bibnamefont {Kahng}},\ and\ \bibinfo
  {author} {\bibfnamefont {S.}~\bibnamefont {Redner}},\ }\bibfield  {title}
  {\bibinfo {title} {Infinite-order percolation and giant fluctuations in a
  protein interaction network},\ }\href
  {https://doi.org/https://doi.org/10.1103/PhysRevE.66.055101} {\bibfield
  {journal} {\bibinfo  {journal} {Physical Review E}\ }\textbf {\bibinfo
  {volume} {66}},\ \bibinfo {pages} {055101} (\bibinfo {year}
  {2002})}\BibitemShut {NoStop}%
\bibitem [{\citenamefont {V{\'a}zquez}(2003)}]{vazquez2003growing}%
  \BibitemOpen
  \bibfield  {author} {\bibinfo {author} {\bibfnamefont {A.}~\bibnamefont
  {V{\'a}zquez}},\ }\bibfield  {title} {\bibinfo {title} {Growing network with
  local rules: Preferential attachment, clustering hierarchy, and degree
  correlations},\ }\href
  {https://doi.org/https://doi.org/10.1103/PhysRevE.67.056104} {\bibfield
  {journal} {\bibinfo  {journal} {Physical Review E}\ }\textbf {\bibinfo
  {volume} {67}},\ \bibinfo {pages} {056104} (\bibinfo {year}
  {2003})}\BibitemShut {NoStop}%
\bibitem [{\citenamefont {Krapivsky}\ and\ \citenamefont
  {Redner}(2005)}]{krapivsky2005network}%
  \BibitemOpen
  \bibfield  {author} {\bibinfo {author} {\bibfnamefont {P.~L.}\ \bibnamefont
  {Krapivsky}}\ and\ \bibinfo {author} {\bibfnamefont {S.}~\bibnamefont
  {Redner}},\ }\bibfield  {title} {\bibinfo {title} {Network growth by
  copying},\ }\href
  {https://doi.org/https://doi.org/10.1103/PhysRevE.71.036118} {\bibfield
  {journal} {\bibinfo  {journal} {Physical Review E}\ }\textbf {\bibinfo
  {volume} {71}},\ \bibinfo {pages} {036118} (\bibinfo {year}
  {2005})}\BibitemShut {NoStop}%
\bibitem [{\citenamefont {Ohno}(1970)}]{ohno1970evolution}%
  \BibitemOpen
  \bibfield  {author} {\bibinfo {author} {\bibfnamefont {S.}~\bibnamefont
  {Ohno}},\ }\href {https://doi.org/10.1007/978-3-642-86659-3} {\emph {\bibinfo
  {title} {Evolution by gene duplication}}}\ (\bibinfo  {publisher}
  {Springer-Verlag},\ \bibinfo {year} {1970})\BibitemShut {NoStop}%
\bibitem [{\citenamefont {Newman}(2018)}]{newman2018networks}%
  \BibitemOpen
  \bibfield  {author} {\bibinfo {author} {\bibfnamefont {M.}~\bibnamefont
  {Newman}},\ }\href@noop {} {\emph {\bibinfo {title} {Networks}}}\ (\bibinfo
  {publisher} {Oxford University Press},\ \bibinfo {year} {2018})\BibitemShut
  {NoStop}%
\bibitem [{\citenamefont {Kumar}\ \emph {et~al.}(2000)\citenamefont {Kumar},
  \citenamefont {Raghavan}, \citenamefont {Rajagopalan}, \citenamefont
  {Sivakumar}, \citenamefont {Tomkins},\ and\ \citenamefont
  {Upfal}}]{kumar2000stochastic}%
  \BibitemOpen
  \bibfield  {author} {\bibinfo {author} {\bibfnamefont {R.}~\bibnamefont
  {Kumar}}, \bibinfo {author} {\bibfnamefont {P.}~\bibnamefont {Raghavan}},
  \bibinfo {author} {\bibfnamefont {S.}~\bibnamefont {Rajagopalan}}, \bibinfo
  {author} {\bibfnamefont {D.}~\bibnamefont {Sivakumar}}, \bibinfo {author}
  {\bibfnamefont {A.}~\bibnamefont {Tomkins}},\ and\ \bibinfo {author}
  {\bibfnamefont {E.}~\bibnamefont {Upfal}},\ }\bibfield  {title} {\bibinfo
  {title} {Stochastic models for the web graph},\ }in\ \href
  {https://doi.org/https://doi.org/10.1109/SFCS.2000.892065} {\emph {\bibinfo
  {booktitle} {Proceedings 41st Annual Symposium on Foundations of Computer
  Science}}}\ (\bibinfo {organization} {IEEE},\ \bibinfo {year} {2000})\ pp.\
  \bibinfo {pages} {57--65}\BibitemShut {NoStop}%
\bibitem [{\citenamefont {Bhat}\ \emph {et~al.}(2016)\citenamefont {Bhat},
  \citenamefont {Krapivsky}, \citenamefont {Lambiotte},\ and\ \citenamefont
  {Redner}}]{bhat2016densification}%
  \BibitemOpen
  \bibfield  {author} {\bibinfo {author} {\bibfnamefont {U.}~\bibnamefont
  {Bhat}}, \bibinfo {author} {\bibfnamefont {P.}~\bibnamefont {Krapivsky}},
  \bibinfo {author} {\bibfnamefont {R.}~\bibnamefont {Lambiotte}},\ and\
  \bibinfo {author} {\bibfnamefont {S.}~\bibnamefont {Redner}},\ }\bibfield
  {title} {\bibinfo {title} {Densification and structural transitions in
  networks that grow by node copying},\ }\href
  {https://doi.org/10.1103/PhysRevE.94.062302} {\bibfield  {journal} {\bibinfo
  {journal} {Physical Review E}\ }\textbf {\bibinfo {volume} {94}},\ \bibinfo
  {pages} {062302} (\bibinfo {year} {2016})}\BibitemShut {NoStop}%
\bibitem [{\citenamefont {Dorogovtsev}\ \emph {et~al.}(2000)\citenamefont
  {Dorogovtsev}, \citenamefont {Mendes},\ and\ \citenamefont
  {Samukhin}}]{dorogovtsev2000structure}%
  \BibitemOpen
  \bibfield  {author} {\bibinfo {author} {\bibfnamefont {S.~N.}\ \bibnamefont
  {Dorogovtsev}}, \bibinfo {author} {\bibfnamefont {J.~F.~F.}\ \bibnamefont
  {Mendes}},\ and\ \bibinfo {author} {\bibfnamefont {A.~N.}\ \bibnamefont
  {Samukhin}},\ }\bibfield  {title} {\bibinfo {title} {Structure of growing
  networks with preferential linking},\ }\href
  {https://doi.org/https://doi.org/10.1103/PhysRevLett.85.4633} {\bibfield
  {journal} {\bibinfo  {journal} {Physical Review Letters}\ }\textbf {\bibinfo
  {volume} {85}},\ \bibinfo {pages} {4633} (\bibinfo {year}
  {2000})}\BibitemShut {NoStop}%
\bibitem [{\citenamefont {Barab{\'a}si}\ \emph {et~al.}(1999)\citenamefont
  {Barab{\'a}si}, \citenamefont {Albert},\ and\ \citenamefont
  {Jeong}}]{barabasi1999mean}%
  \BibitemOpen
  \bibfield  {author} {\bibinfo {author} {\bibfnamefont {A.-L.}\ \bibnamefont
  {Barab{\'a}si}}, \bibinfo {author} {\bibfnamefont {R.}~\bibnamefont
  {Albert}},\ and\ \bibinfo {author} {\bibfnamefont {H.}~\bibnamefont
  {Jeong}},\ }\bibfield  {title} {\bibinfo {title} {Mean-field theory for
  scale-free random networks},\ }\href
  {https://doi.org/https://doi.org/10.1016/S0378-4371(99)00291-5} {\bibfield
  {journal} {\bibinfo  {journal} {Physica A: Statistical Mechanics and its
  Applications}\ }\textbf {\bibinfo {volume} {272}},\ \bibinfo {pages} {173}
  (\bibinfo {year} {1999})}\BibitemShut {NoStop}%
\bibitem [{\citenamefont {Dorogovtsev}\ and\ \citenamefont
  {Mendes}(2022)}]{dorogovtsev2022nature}%
  \BibitemOpen
  \bibfield  {author} {\bibinfo {author} {\bibfnamefont {S.~N.}\ \bibnamefont
  {Dorogovtsev}}\ and\ \bibinfo {author} {\bibfnamefont {F.}~\bibnamefont
  {Mendes}},\ }\href@noop {} {\emph {\bibinfo {title} {The nature of complex
  networks}}}\ (\bibinfo  {publisher} {Oxford University Press},\ \bibinfo
  {year} {2022})\BibitemShut {NoStop}%
\bibitem [{\citenamefont {Ispolatov}\ \emph
  {et~al.}(2005{\natexlab{a}})\citenamefont {Ispolatov}, \citenamefont
  {Krapivsky},\ and\ \citenamefont {Yuryev}}]{ispolatov2005duplication}%
  \BibitemOpen
  \bibfield  {author} {\bibinfo {author} {\bibfnamefont {I.}~\bibnamefont
  {Ispolatov}}, \bibinfo {author} {\bibfnamefont {P.~L.}\ \bibnamefont
  {Krapivsky}},\ and\ \bibinfo {author} {\bibfnamefont {A.}~\bibnamefont
  {Yuryev}},\ }\bibfield  {title} {\bibinfo {title} {Duplication-divergence
  model of protein interaction network},\ }\href
  {https://doi.org/10.1103/PhysRevE.71.061911} {\bibfield  {journal} {\bibinfo
  {journal} {Physical Review E}\ }\textbf {\bibinfo {volume} {71}},\ \bibinfo
  {pages} {061911} (\bibinfo {year} {2005}{\natexlab{a}})}\BibitemShut
  {NoStop}%
\bibitem [{\citenamefont {Cai}\ \emph {et~al.}(2015)\citenamefont {Cai},
  \citenamefont {Liu},\ and\ \citenamefont {Lee}}]{cai2015mean}%
  \BibitemOpen
  \bibfield  {author} {\bibinfo {author} {\bibfnamefont {S.}~\bibnamefont
  {Cai}}, \bibinfo {author} {\bibfnamefont {Z.}~\bibnamefont {Liu}},\ and\
  \bibinfo {author} {\bibfnamefont {H.}~\bibnamefont {Lee}},\ }\bibfield
  {title} {\bibinfo {title} {Mean field theory for biology inspired
  duplication-divergence network model},\ }\bibfield  {journal} {\bibinfo
  {journal} {Chaos: An Interdisciplinary Journal of Nonlinear Science}\
  }\textbf {\bibinfo {volume} {25}},\ \href
  {https://doi.org/https://doi.org/10.1063/1.4928212}
  {https://doi.org/10.1063/1.4928212} (\bibinfo {year} {2015})\BibitemShut
  {NoStop}%
\bibitem [{\citenamefont {Farid}\ and\ \citenamefont
  {Christensen}(2006)}]{farid2006evolving}%
  \BibitemOpen
  \bibfield  {author} {\bibinfo {author} {\bibfnamefont {N.}~\bibnamefont
  {Farid}}\ and\ \bibinfo {author} {\bibfnamefont {K.}~\bibnamefont
  {Christensen}},\ }\bibfield  {title} {\bibinfo {title} {Evolving networks
  through deletion and duplication},\ }\href
  {https://doi.org/10.1088/1367-2630/8/9/212} {\bibfield  {journal} {\bibinfo
  {journal} {New Journal of Physics}\ }\textbf {\bibinfo {volume} {8}},\
  \bibinfo {pages} {212} (\bibinfo {year} {2006})}\BibitemShut {NoStop}%
\bibitem [{\citenamefont {Pastor-Satorras}\ \emph {et~al.}(2003)\citenamefont
  {Pastor-Satorras}, \citenamefont {Smith},\ and\ \citenamefont
  {Sol{\'e}}}]{pastor2003evolving}%
  \BibitemOpen
  \bibfield  {author} {\bibinfo {author} {\bibfnamefont {R.}~\bibnamefont
  {Pastor-Satorras}}, \bibinfo {author} {\bibfnamefont {E.}~\bibnamefont
  {Smith}},\ and\ \bibinfo {author} {\bibfnamefont {R.~V.}\ \bibnamefont
  {Sol{\'e}}},\ }\bibfield  {title} {\bibinfo {title} {Evolving protein
  interaction networks through gene duplication},\ }\href
  {https://doi.org/10.1016/S0022-5193(03)00028-6} {\bibfield  {journal}
  {\bibinfo  {journal} {Journal of Theoretical Biololgy}\ }\textbf {\bibinfo
  {volume} {222}},\ \bibinfo {pages} {199} (\bibinfo {year}
  {2003})}\BibitemShut {NoStop}%
\bibitem [{\citenamefont {Schneider}\ \emph {et~al.}(2011)\citenamefont
  {Schneider}, \citenamefont {de~Arcangelis},\ and\ \citenamefont
  {Herrmann}}]{schneider2011modeling}%
  \BibitemOpen
  \bibfield  {author} {\bibinfo {author} {\bibfnamefont {C.~M.}\ \bibnamefont
  {Schneider}}, \bibinfo {author} {\bibfnamefont {L.}~\bibnamefont
  {de~Arcangelis}},\ and\ \bibinfo {author} {\bibfnamefont {H.~J.}\
  \bibnamefont {Herrmann}},\ }\bibfield  {title} {\bibinfo {title} {Modeling
  the topology of protein interaction networks},\ }\href
  {https://doi.org/https://doi.org/10.1103/PhysRevE.84.016112} {\bibfield
  {journal} {\bibinfo  {journal} {Physical Review E}\ }\textbf {\bibinfo
  {volume} {84}},\ \bibinfo {pages} {016112} (\bibinfo {year}
  {2011})}\BibitemShut {NoStop}%
\bibitem [{\citenamefont {V{\'a}zquez}\ \emph {et~al.}(2003)\citenamefont
  {V{\'a}zquez}, \citenamefont {Flammini}, \citenamefont {Maritan},\ and\
  \citenamefont {Vespignani}}]{vazquez2003modeling}%
  \BibitemOpen
  \bibfield  {author} {\bibinfo {author} {\bibfnamefont {A.}~\bibnamefont
  {V{\'a}zquez}}, \bibinfo {author} {\bibfnamefont {A.}~\bibnamefont
  {Flammini}}, \bibinfo {author} {\bibfnamefont {A.}~\bibnamefont {Maritan}},\
  and\ \bibinfo {author} {\bibfnamefont {A.}~\bibnamefont {Vespignani}},\
  }\bibfield  {title} {\bibinfo {title} {Modeling of protein interaction
  networks},\ }\href {https://doi.org/10.1017/CBO9780511845086.006} {\bibfield
  {journal} {\bibinfo  {journal} {Complexus}\ }\textbf {\bibinfo {volume}
  {1}},\ \bibinfo {pages} {38} (\bibinfo {year} {2003})}\BibitemShut {NoStop}%
\bibitem [{\citenamefont {Ispolatov}\ \emph
  {et~al.}(2005{\natexlab{b}})\citenamefont {Ispolatov}, \citenamefont
  {Krapivsky}, \citenamefont {Mazo},\ and\ \citenamefont
  {Yuryev}}]{ispolatov2005cliques}%
  \BibitemOpen
  \bibfield  {author} {\bibinfo {author} {\bibfnamefont {I.}~\bibnamefont
  {Ispolatov}}, \bibinfo {author} {\bibfnamefont {P.}~\bibnamefont
  {Krapivsky}}, \bibinfo {author} {\bibfnamefont {I.}~\bibnamefont {Mazo}},\
  and\ \bibinfo {author} {\bibfnamefont {A.}~\bibnamefont {Yuryev}},\
  }\bibfield  {title} {\bibinfo {title} {Cliques and duplication--divergence
  network growth},\ }\href {https://doi.org/10.1088/1367-2630/7/1/145}
  {\bibfield  {journal} {\bibinfo  {journal} {New Journal of Physics}\ }\textbf
  {\bibinfo {volume} {7}},\ \bibinfo {pages} {145} (\bibinfo {year}
  {2005}{\natexlab{b}})}\BibitemShut {NoStop}%
\bibitem [{\citenamefont {Sudbrack}\ \emph {et~al.}(2018)\citenamefont
  {Sudbrack}, \citenamefont {Brunnet}, \citenamefont {de~Almeida},
  \citenamefont {Ferreira},\ and\ \citenamefont
  {Gamermann}}]{sudbrack2018master}%
  \BibitemOpen
  \bibfield  {author} {\bibinfo {author} {\bibfnamefont {V.}~\bibnamefont
  {Sudbrack}}, \bibinfo {author} {\bibfnamefont {L.~G.}\ \bibnamefont
  {Brunnet}}, \bibinfo {author} {\bibfnamefont {R.~M.}\ \bibnamefont
  {de~Almeida}}, \bibinfo {author} {\bibfnamefont {R.~M.}\ \bibnamefont
  {Ferreira}},\ and\ \bibinfo {author} {\bibfnamefont {D.}~\bibnamefont
  {Gamermann}},\ }\bibfield  {title} {\bibinfo {title} {Master equation for the
  degree distribution of a duplication and divergence network},\ }\href
  {https://doi.org/https://doi.org/10.1016/j.physa.2018.06.066} {\bibfield
  {journal} {\bibinfo  {journal} {Physica A: Statistical Mechanics and its
  Applications}\ }\textbf {\bibinfo {volume} {509}},\ \bibinfo {pages} {588}
  (\bibinfo {year} {2018})}\BibitemShut {NoStop}%
\bibitem [{\citenamefont {Newman}(2007)}]{newman2007component}%
  \BibitemOpen
  \bibfield  {author} {\bibinfo {author} {\bibfnamefont {M.~E.}\ \bibnamefont
  {Newman}},\ }\bibfield  {title} {\bibinfo {title} {Component sizes in
  networks with arbitrary degree distributions},\ }\href
  {https://doi.org/https://doi.org/10.1103/PhysRevE.76.045101} {\bibfield
  {journal} {\bibinfo  {journal} {Physical Review E}\ }\textbf {\bibinfo
  {volume} {76}},\ \bibinfo {pages} {045101} (\bibinfo {year}
  {2007})}\BibitemShut {NoStop}%
\bibitem [{\citenamefont {Coniglio}(1982)}]{coniglio1982cluster}%
  \BibitemOpen
  \bibfield  {author} {\bibinfo {author} {\bibfnamefont {A.}~\bibnamefont
  {Coniglio}},\ }\bibfield  {title} {\bibinfo {title} {Cluster structure near
  the percolation threshold},\ }\href
  {https://doi.org/https://10.1088/0305-4470/15/12/032} {\bibfield  {journal}
  {\bibinfo  {journal} {Journal of Physics A: Mathematical and General}\
  }\textbf {\bibinfo {volume} {15}},\ \bibinfo {pages} {3829} (\bibinfo {year}
  {1982})}\BibitemShut {NoStop}%
\bibitem [{\citenamefont {Sol{\'e}}\ and\ \citenamefont
  {Valverde}(2008)}]{sole2008spontaneous}%
  \BibitemOpen
  \bibfield  {author} {\bibinfo {author} {\bibfnamefont {R.~V.}\ \bibnamefont
  {Sol{\'e}}}\ and\ \bibinfo {author} {\bibfnamefont {S.}~\bibnamefont
  {Valverde}},\ }\bibfield  {title} {\bibinfo {title} {Spontaneous emergence of
  modularity in cellular networks},\ }\href
  {https://doi.org/10.1098/rsif.2007.1108} {\bibfield  {journal} {\bibinfo
  {journal} {Journal of Royal Society Interface}\ }\textbf {\bibinfo {volume}
  {5}},\ \bibinfo {pages} {129} (\bibinfo {year} {2008})}\BibitemShut {NoStop}%
\bibitem [{Note1()}]{Note1}%
  \BibitemOpen
  \bibinfo {note} {This process may remind a single-gene duplication evolution;
  multiple genes duplications events (e.g., entire genome duplication) may be
  possible but they are not considered here in favor of a minimal
  approach.}\BibitemShut {Stop}%
\bibitem [{\citenamefont {Force}\ \emph {et~al.}(1999)\citenamefont {Force},
  \citenamefont {Lynch}, \citenamefont {Pickett}, \citenamefont {Amores},
  \citenamefont {Yan},\ and\ \citenamefont
  {Postlethwait}}]{force1999preservation}%
  \BibitemOpen
  \bibfield  {author} {\bibinfo {author} {\bibfnamefont {A.}~\bibnamefont
  {Force}}, \bibinfo {author} {\bibfnamefont {M.}~\bibnamefont {Lynch}},
  \bibinfo {author} {\bibfnamefont {F.~B.}\ \bibnamefont {Pickett}}, \bibinfo
  {author} {\bibfnamefont {A.}~\bibnamefont {Amores}}, \bibinfo {author}
  {\bibfnamefont {Y.-l.}\ \bibnamefont {Yan}},\ and\ \bibinfo {author}
  {\bibfnamefont {J.}~\bibnamefont {Postlethwait}},\ }\bibfield  {title}
  {\bibinfo {title} {Preservation of duplicate genes by complementary,
  degenerative mutations},\ }\href
  {https://doi.org/https://doi.org/10.1093/genetics/151.4.1531} {\bibfield
  {journal} {\bibinfo  {journal} {Genetics}\ }\textbf {\bibinfo {volume}
  {151}},\ \bibinfo {pages} {1531} (\bibinfo {year} {1999})}\BibitemShut
  {NoStop}%
\bibitem [{\citenamefont {Krapivsky}\ \emph {et~al.}(2010)\citenamefont
  {Krapivsky}, \citenamefont {Redner},\ and\ \citenamefont
  {Ben-Naim}}]{krapivsky2010kinetic}%
  \BibitemOpen
  \bibfield  {author} {\bibinfo {author} {\bibfnamefont {P.~L.}\ \bibnamefont
  {Krapivsky}}, \bibinfo {author} {\bibfnamefont {S.}~\bibnamefont {Redner}},\
  and\ \bibinfo {author} {\bibfnamefont {E.}~\bibnamefont {Ben-Naim}},\
  }\href@noop {} {\emph {\bibinfo {title} {A Kinetic View of Statistical
  Physics}}}\ (\bibinfo  {publisher} {Cambridge University Press},\ \bibinfo
  {year} {2010})\BibitemShut {NoStop}%
\bibitem [{\citenamefont {Dorogovtsev}\ \emph {et~al.}(2002)\citenamefont
  {Dorogovtsev}, \citenamefont {Mendes},\ and\ \citenamefont
  {Samukhin}}]{dorogovtsev2002multifractal}%
  \BibitemOpen
  \bibfield  {author} {\bibinfo {author} {\bibfnamefont {S.~N.}\ \bibnamefont
  {Dorogovtsev}}, \bibinfo {author} {\bibfnamefont {J.~F.~F.}\ \bibnamefont
  {Mendes}},\ and\ \bibinfo {author} {\bibfnamefont {A.}~\bibnamefont
  {Samukhin}},\ }\bibfield  {title} {\bibinfo {title} {Multifractal properties
  of growing networks},\ }\href
  {https://doi.org/https://10.1209/epl/i2002-00465-1} {\bibfield  {journal}
  {\bibinfo  {journal} {Europhysics Letters}\ }\textbf {\bibinfo {volume}
  {57}},\ \bibinfo {pages} {334} (\bibinfo {year} {2002})}\BibitemShut
  {NoStop}%
\bibitem [{\citenamefont {Onsager}(1949)}]{onsager1949statistical}%
  \BibitemOpen
  \bibfield  {author} {\bibinfo {author} {\bibfnamefont {L.}~\bibnamefont
  {Onsager}},\ }\bibfield  {title} {\bibinfo {title} {Statistical
  hydrodynamics},\ }\href@noop {} {\bibfield  {journal} {\bibinfo  {journal}
  {Il Nuovo Cimento (1943-1954)}\ }\textbf {\bibinfo {volume} {6}},\ \bibinfo
  {pages} {279} (\bibinfo {year} {1949})}\BibitemShut {NoStop}%
\end{thebibliography}%
\appendix
\vspace{-0.15in}
\section{Number of edges $\langle E_t \rangle$ for $\alpha\neq 0, \beta = 0$}
\label{app_A}

A recurrence for $\langle E_t \rangle$ has the following form

\begin{equation}
	\langle E_{t+1} \rangle - \langle E_{t} \rangle = 2(1-\delta) \frac{\langle E_t \rangle}{t}  + \alpha.
\label{E_t1_AppA}
\end{equation}

A continuum approximation recasts it as

\begin{equation}
	\frac{d\langle E_t \rangle }{dt} = 2(1-\delta) \frac{\langle E_t \rangle}{t}+ \alpha.
	\label{E_t2_AppA}
\end{equation}
	
Solving (\ref{E_t2_AppA}) we get 

\begin{equation}
\langle E_t \rangle  \sim 
\begin{cases}
\frac{\alpha}{2\delta -1}t + C_{t0}t^{2(1-\delta)}, \hspace{0.52cm}\mathrm{for \; \;} \delta \gtrless 1/2,\\
\alpha t \mathrm{ln}(t) + C_{t_0}t, \hspace{1.2cm}\mathrm{for \; \;}\delta=1/2,
\label{E_t3_AppA}
\end{cases}
\end{equation}

$C_{t_0}$ an integration constant. The scaling with $t$ of the mean vertex degree $\langle k_t \rangle$ follows directly from $2\langle E_t \rangle/t$.

\vspace{-0.15in}
\section{Number of edges $\langle E_t \rangle$ for $\alpha\neq 0, \beta \neq 0$}
\label{app_B}

Here, the recurrence for $\langle E_t \rangle$ is

\begin{equation}
\begin{aligned}
	\langle E_{t+1} \rangle - \langle E_{t} \rangle =  (1-\delta) \frac{2\langle E_t \rangle}{t} + \alpha + \beta \left( t -\frac{2\langle E_t \rangle}{t} -1 \right),
\end{aligned}
\label{E_t1_AppB}
\end{equation}

which is written in a continuum form as 

\begin{equation}
\begin{aligned}
	& \frac{d\langle E_t \rangle}{dt}=  2(1-\delta) \frac{\langle E_t \rangle}{t}  &+ \alpha + \beta \left( t - 2\frac{\langle E_t \rangle}{t} -1 \right).
\end{aligned}
\label{E_t2_AppB}
\end{equation}

A solution of (\ref{E_t2_AppB}), for $2\beta \neq 1 - 2\delta$, $\delta >0$, gives

\begin{equation}
	\langle E_t \rangle \sim \frac{\beta}{2(\delta + \beta)}t^2 - \frac{\beta - \alpha}{2(\delta + \beta)-1}t + C_{t_0}t^{2(1-\delta -\beta)},
\end{equation}

and, for $2\beta = 1-2\delta$

\begin{equation}
	\langle E_t \rangle \sim t^2 \left( \frac{1}{2} - \delta \right) + t\mathrm{ln(t)} \left( \delta + \alpha - \frac{1}{2} \right) + C_{t_0} t.
\end{equation}

with $C_{t_0}$ an integration constant.
\end{document}